\documentclass[twocolumn,showpacs,preprintnumbers,amsmath,amssymb]{revtex4}

\usepackage{graphics}
\usepackage{epsfig}
\usepackage{dcolumn}
\usepackage{bm}

\begin{document}




\title{Comment on ``Chiral Suppression of Scalar Glueball Decay''}

\author{Kuang-Ta Chao$^1$, Xiao-Gang He$^{2}$, and Jian-Ping Ma$^{3,1}$}


\affiliation{$^1$Department of Physics, Peking University,
Beijing\\
$^2$Department of Physics and Center for Theoretical Sciences,
National Taiwan University,
Taipei\\
$^3$Institute Of Theoretical Physics, Academia Sinica, Beijing}


\par
\maketitle

In a recent letter, based on an effective Lagrangian,
Chanowitz\cite{Chan} showed that in the limit that the mass $m_q$
of a light quark $q$ goes to zero, the decay amplitude for a
scalar glueball $G_s$ decaying into $q \bar q$ goes to zero, and
conjectured further that this chiral suppression also occurs at
the hadron level for $G_s$ decays into $\pi\pi, KK$ with the ratio
of these two branching ratios to be of the order $\mathcal
O(m^2_{u,d}/m^2_s)$ for finite quark masses. Here we show that the
decay $G_s \to q\bar q$ is forbidden in the chiral limit in QCD
without assumptions.  More essentially, we show that this chiral
suppression  may be spoiled and may not materialize itself at the
hadron level.

A glueball here is assumed to be a pure gluonic state. It decays
into a $q\bar q$ pair through a multi-gluon annihilation process.
The decay amplitude for $G_s \to q(p_1) \bar q(p_2)$ can be
written as a product of a spinor pair $\bar u(p_1) $ and $v(p_2)$
with a product of any number of $\gamma$ matrices sandwiched
between the spinors. Because vector-like coupling in QCD, for
$m_q=0$ the number of the $\gamma$-matrices is an odd number which
can always be reduced to one $\gamma$-matrix. Therefore the
amplitude can be written as:
\begin{equation}
{\mathcal T}_{q\bar q} = \bar u(p_1) \gamma_{\mu} A^\mu v(p_2).
\nonumber
\end{equation}
Lorentz covariance of the amplitude then dictates $A^\mu
(p_1,p_2)$ to be of the form $a_1 p_1^\mu + a_2 p_2^\mu$.
Therefore in the chiral limit $m_q =0$, ${\mathcal T}_{q \bar q} =
0$. The result also applies to a pseudoscalar glueball decays into
a $q \bar q$ pair.

To study whether there is a chiral suppression in $G_s \to \pi\pi,
KK$ or not, we work with an effective Lagrangian, $L_s = f_g
G^{a,\mu\nu} G^a_{\ \mu\nu} G_s$, as in \cite{Chan}, and employ
QCD factorization\cite{BrLe} to calculate the amplitude $\mathcal
{T}_{\pi\pi}$ for $G_s\to \pi^+\pi^-$.
To the leading twist-2 order, there are two diagrams
with  the two gluons splitting into two quarks and two
anti-quarks, and then form two pions. The two gluons are off-shell
by the scale at order of $M_{G_s}$. A direct calculation gives:
\begin{eqnarray}
&&{\mathcal T}_{\pi\pi} = - \alpha_s  f_g \frac{ 8\pi}{9} f_\pi^2
\int_0^1 du_1 du_2 \phi_{\pi^+} (u_1) \phi_{\pi^-} (u_2)
\nonumber\\
&&\times \left ( \frac{1}{u_1 u_2} + \frac{1}{(1-u_1)(1-u_2)}
\right )\left [ 1 + {\mathcal O}
(\alpha_s,\lambda/M_{G_s})\right],\nonumber
\end{eqnarray}
where $\phi_\pi$ is normalized as $\int d u \phi_\pi (u) =1$.
$u_i(i=1,2)$ is the momentum fraction carried by the anti-quark in
the meson. In the above, $\lambda$ can be any soft scale, such as
quark mass, $\Lambda_{QCD}$ and $m_\pi$. Clearly, ${\mathcal T}_{\pi\pi}$
is not zero in the chiral limit $m_q=0$.
\par
The amplitude for $G_s \to K^+K^-$ decay can be obtained by
replacing quantities related to  $\pi$ by those related to $K$
correspondingly. We would obtain, $R={B(G_s \to \pi\pi)/ B(G_s \to
KK)} \approx {f_\pi^4/f_K^4} = 0.48$, which is substantially
different from 1. This suppression is much milder compared with
the one at the quark level. This is due to the fact that in
perturbative QCD (pQCD) calculation the decay of $G_s \to \pi \pi,
KK$ is related to the coupling of $G_s$ to two pairs of $q \bar q$
compared with conjectured by Chanowitz in \cite{Chan}, where it is
assumed that $G_s$ just couples to one $q \bar q$ pair. We should
point out that whether the chiral suppression at quark level can
be realized still waits for better non-perturbative calculation
for the direct two quark hadronization into $\pi\pi$ and $K K$. If
the pQCD contribution dominates, the result of $R \approx
{f_\pi^4/f_K^4}$ can be obtained without the assumption of the
effective Lagrangian. Because glueball is a pure gluon state, the
amplitude of the decay $G_s\to \pi^+\pi^-$ can always be written
with QCD factorization as $T_{\pi\pi}= f_\pi^2 H_g \otimes
\phi_{\pi^+} \otimes \phi_{\pi^-}$, where the higher-twist effects
related to $\pi$'s are neglected and $H_g$ consists of some
perturbative coefficient functions and some quantities related to
the structure of $G_s$. Although $H_g$ is unknown, one can easily
find the result of $R \approx {f_\pi^4/f_K^4}$.

The $f_0(1710)$ is a candidate for scalar glueball. Early
measurement obtained $R \leq 0.11$\cite{PDG}, and a larger one by
BES\cite{bes} $R=0.41^{+0.11}_{-0.17}$ recently. It is interesting
to notice that the later is consistent with our result and may
favor that the $f_0(1710)$ is a gluebal. However one should
remember that the prediction $R \approx {f_\pi^4/f_K^4}$ can have
substantial non-perturbative corrections and  there may be further
complication by mixing effects of a glueball with $q\bar q$
states. A more detailed study can be found in \cite{CHM}.

\noindent {\bf Acknowledgments:} This work was supported in part
by grants from NSC and NNSFC (No 10421503).

\end{document}